\documentclass[twocolumn,prl,aps]{revtex4}

\begin{document}

{\bf Comment on ``Dynamic behavior of anisotropic non-equilibrium
driving lattice gases''}

In a recent Letter Albano and Saracco (AS)~\cite{as-02}  study the dynamic
behavior of some anisotropic driven lattice gases by
Monte Carlo (MC) simulations~\cite{STMC}.
They check the scaling of the order parameter $OP$ defined in their Eq.~(4)
and, by data collapse, determine the relevant scaling exponents. Then,
they formulate scaling Ans\"atze (see Eqs.~(5) and (8) 
of AS) in order to relate the measured scaling exponents 
with the critical exponents analytically computed within different 
field-theoretical approaches 
(see Refs.~\cite{js-lc-86,Garrido98}). 
Comparing MC results and theoretical predictions
they conclude that the universality class of the phase 
transition in the driven lattice gases considered in the paper 
is well-described by the field theory proposed in
Ref.~\cite{Garrido98}, ruling out that of Ref.~\cite{js-lc-86}.
In this Comment we point out that the 
Ans\"atze (5) and (8) 
do not take properly into account the strongly anisotropic nature of
the phase transition, by implicitly assuming $z = z_{\bot} = z_{\parallel}$. 
As a consequence, at variance with 
AS's claims, MC data are not conclusive on the issue of the  
universality class.

Let us outline the argument. 
During the relaxation of a spin system from an initial 
uncorrelated state with small magnetization $m_0$, the 
$k$-th moment of the magnetization is expected to scale according 
to~\cite{STMC}
$
M^{(k)}(t,\tau, L, m_0) = 
b^{-k\beta/\nu}f_k(b^{-z}t,b^{1/\nu},b^{1/\nu}\tau,b^{-1}L, b^{x_0}m_0)
$,
where $x_0$ is related to the initial-slip exponent.
The generalization of this expression to a system that is strongly anisotropic 
and has a conserved order parameter requires some care.
Using the notations of Ref.~\cite{sz-95}, one expects
$
M^{(k)}(t,\tau, L_\parallel,L_\bot, m_0) = 
b^{-k\beta/\nu_\parallel}
 f_k(b^{-z_\parallel}t,b^{1/\nu_\parallel}\tau,b^{-1}L_\parallel, 
   b^{-\nu_\bot/\nu_\parallel}L_\bot, b^{x_0}m_0)
$,
where $M$ can be naturally identified with some suitable order parameter,
for instance with the excess density as in AS. 
Since $m_0$ is approximately zero for the initial configurations, 
taking into account that the $OP$ defined by AS is expected
to scale as $\sqrt{M^{(2)}}$~\cite{as-02}, 
one obtains Eq. (5) of AS with the identification $z = z_\parallel$. 
Correctly, in Table I, AS report the predictions $z = 4/3$ and 
$z \approx 1.998$ for the two models they consider (see Eqs. (1) and (2) of AS),
which refer indeed to $z_\parallel$. Formulae 
(6) and (7) follow from Eq. (5) and therefore they are correct only if 
we identify $z = z_\parallel$. 
In Eq.~(8), AS generalize the scaling behavior of $OP$ to
the case of the relaxation 
from the ground state configuration.
Without attemping a derivation of their equation, we simply note
that, as it stands, it is not compatible with Eq.~(5) of AS:
The exponent $z$ appearing in Eq.~(8) must be identified with 
$z_\bot = \nu_\parallel z_\parallel/\nu_\bot$ and not with $z_\parallel$.
Indeed, whatever the initial condition is, we expect that   
after a time interval long enough to allow the relaxation of the initial
condition, $OP$ will display a characteristic long-time behavior,
that should be captured by any reasonable scaling form.
Now, from Eq.~(5) (rewritten as before) 
it is easy to derive that, 
in the thermodynamic limit,
$OP(t,\tau) = t^{-\beta/\nu_\parallel z_\parallel}f_{\rm
FD}(t^{1/\nu_\parallel z_\parallel}\tau)$, while from Eq.~(8)
$OP(t,\tau) = t^{-\beta/\nu_\bot z}f_{\rm
GS}(t^{1/\nu_\bot z}\tau)$.
These two 
equations give the same long-time limit only if in the second case 
we identify $z$ with $z_\bot$, i.e.
$
OP(t,\tau,L_\parallel,L_\bot) = b^{-\beta/\nu_\bot} 
OP(b^{-z_\bot} 
t, b^{1/\nu_\bot}\tau, b^{-\nu_\parallel/\nu_\bot}L_\parallel, 
b^{-1}L_\bot) 
$.
Equations (9) and (10) follow from Eq.~(8) and therefore 
here $z$ must be interpreted as $z_\bot$.
As a consequence, some of the theoretical predictions for the
quantities measured by AS and reported in Tab.~I of Ref.~\cite{as-02},
change according to the following table 
\begin{center}
\begin{tabular}{ccc}
\hline
Model &
\( \beta/\nu_{\bot} z (+) \)&
\( c_{\bot} (+) \)\\
\hline
Eq. (1) in Ref.~\cite{as-02}&
$1/4 $ &
$1/4 $ \\
Eq. (2) in Ref.~\cite{as-02}&
$ \approx 0.13 $&
$ \approx 0.27 $\\
\hline
\end{tabular}
\end{center}
Comparing these corrected predictions with the numerical results 
reported in Tab.~I of AS,
we conclude that the value of $\beta/\nu_{\bot} z$ is in agreement with 
Eq.~(1), which is exactly the opposite conclusion with respect to AS, while
the numerical estimate of $c_{\bot}$ differs from the predictions of 
both field-theoretical models considered in AS's paper.

\vspace{0.5cm}
\noindent
Sergio Caracciolo,${}^{1}$ Andrea  Gambassi,${}^{2}$ Massimiliano
Gubinelli,${}^{3}$ and Andrea Pelissetto${}^{4}$\\
\noindent
{\small ${}^1$~Dipartimento di Fisica and INFN, Universit\`a di Milano, and
INFM-NEST, I-20133 Milano, Italy\\
${}^2$~Max-Planck-Institut f\"ur Metallforschung
and 
Institut f\"ur Theoretische und Angewandte Physik der Universit\"at
Stuttgart, 
D-70569 Stuttgart, Germany\\
${}^3$~Dipartimento di Matematica Applicata and INFN, Universit\`a di
Pisa, I-56100 Pisa, Italy\\
${}^4$~Dipartimento di Fisica and INFN, Universit\`a di Roma ``La
Sapienza'', I-00185 Roma, Italy\\ }


\end{document}